\documentclass[useAMS,usenatbib]{mn2e}
\usepackage{epsfig}
\usepackage{color} 
\usepackage{natbib}
\usepackage{deluxetable}
\usepackage{amssymb}
\usepackage{rotating}
\usepackage{graphicx}
\usepackage[colorlinks=true,citecolor=blue]{hyperref}

\title[Intranight optical variability of radio-quiet WLQs]
{Intranight Optical Variability of Radio-Quiet Weak Emission Line Quasars-II}
\author[$Chand$, $Kumar$ \& $Gopal-Krishna$ ]{Hum Chand$^{1}$\thanks{E-mail: hum@aries.res.in(HC);
parveen@aries.res.in(PK); gopaltani@gmail.com(G-K)}, Parveen Kumar$^{1}$$^{\star}$,
Gopal-Krishna$^{2}$$^{\star}$ \\\\ $^{1}$Aryabhatta Research Institute of Observational Sciences (ARIES),
Manora Peak, Nainital $-$ 263002, India\\ $^{2}$Inter-University
Centre for Astronomy and Astrophysics (IUCAA), Postbag 4, Ganeshkhind, Pune 411 007, India\\}

\begin{document}
\date{Accepted ---. Received ---; in original form ---}

\pagerange{\pageref{firstpage}--\pageref{lastpage}} \pubyear{2014}

\maketitle

\label{firstpage}
\begin{abstract}

This is continuation of our search for the elusive radio-quiet blazars, by
carrying out a systematic programme to detect intranight optical variability
(INOV) in a subset of `Weak-Lines-Quasars' (WLQs) which are designated as
`high confidence BL Lac candidates' and are known to be radio-quiet. For 10
such RQWLQs, we present here the INOV observations taken in 16 sessions of
durations $\ga$ 3.5 hours each. Combining these data with our previously
published INOV monitoring of RQWLQs in 13 sessions, gives a set of INOV
observations of 15 RQWLQs monitored in 29 sessions each lasting more than 3.5
hours. The 29 differential light curves (DLCs), thus obtained for the 15
RQWLQs, were subjected to an statistical analysis using the F$-$test and the
deduced INOV characteristics of the RQWLQs are compared with those
published recently for several prominent AGN classes, also using the
F$-$test. However, since the RQWLQs are generally 1$-$2 magnitudes fainter, a
rigorous comparison has to wait for somewhat more sensitive INOV
observations than those presented
here. Based on our existing INOV observations, it seems that RQWLQs
in our sample show a significantly higher INOV duty cycle than radio-quiet quasars and
radio lobe-dominated quasars. Two sessions when we detected rather strong
(blazar-like) INOV for RQWLQs are pointed out and both these RQWLQs are
therefore candidates for radio-quiet BL Lacs.

\end{abstract}



\section{Introduction} 

The presence of prominent broad emission lines in the optical/UV
spectrum is a hallmark of the Active Galactic Nuclei (AGN) designated
as quasars. However, such lines can appear much weaker for a class of
AGN, called blazars, in which the optical/UV emission is dominated by
the Doppler boosted nonthermal continuum from the relativistic jet and
is therefore substantially polarized. Specifically, this `weak line'
characterization holds for a subclass of blazars, called BL Lac
objects (BLOs), in contrast to the other blazar subclass, called
`highly polarized quasars' (HPQs) which display the emission lines at
a fairly strong level~\citep[e.g.,][and references
  therein]{Urry1995PASP..107..803U}. Being jet dominated, both blazar
subclasses, HPQs and BLOs are radio loud in the sense that the
radio-to-optical flux density ratio $R > 10$, where the radio and
optical continuum flux densities refer to the rest-frame wavelengths
of 6 cm and 2500\AA, respectively
~\citep[e.g.,][]{Kellermann1989AJ.....98.1195K,Stocke1992ApJ...396..487S}.
But, whereas HPQs have an abundant population of weakly polarized,
radio-quiet counterparts (radio-quiet quasars: RQQs), the existence of
radio-quiet analogs of BLOs (RQBLOs) continues to be an open
question.

The large optical survey SDSS ~\citep{York2000AJ....120.1579Y} was used by~\citet{Collinge2005AJ....129.2542C} 
and ~\citet{Anderson2007AJ....133..313A} to find candidates for
radio-quiet BLOs. They termed such candidates ``Weak-Lines-Quasars''
(WLQs). In this way, dozens of WLQs marked by abnormally weak broad
emission-lines~\citep[i.e, rest-frame EW$ < 15.4$\AA~ for the Ly$+$NV
emission-line complex,][]{Diamond-Stanic2009ApJ...699..782D} have been reported
in the literature as summarized in our first paper of this series
~\citep*[][hereafter Paper I]{Gopal2013MNRAS.430.1302G}.
Since many of the WLQs are indeed found to be
radio-quiet ~\citep[e.g.,][]{Plotkin2010AJ....139..390P}, they could potentially
qualify as the elusive RQBLOs. However, they are generally regarded as
weak-lined analogs of RQQs because, in contrast to BLOs (and much like
RQQs), RQWLQs exhibit low optical polarization
~\citep{Smith2007ApJ...663..118S} and mild optical continuum variability on time
scales ranging from days to years~\citep{Plotkin2010ApJ...721..562P}. This is
further corroborated by the similarity observed between the
UV-optical spectral indices, $\alpha$ , of WLQs and RQQs. For RQQs the
median value of $\alpha$ is $-0.52$ as against $-1.15$ for BLO candidates
~\citep{Diamond-Stanic2009ApJ...699..782D,Plotkin2010AJ....139..390P}.

Clearly, the above interpretation of RQWLQs does allow for the
possibility that a small subset of them may indeed be the long-sought
RQBLOs where the optical continuum is significantly, if not
predominantly, contributed by a Doppler boosted relativistic jet. A
potentially fruitful approach to explore this possibility was employed
in Paper I, where we reported the first search for intranight optical
variability (INOV) of RQWLQs. This was motivated by the well
established result that BLOs exhibit a distinctly stronger INOV, both
in amplitude ($\psi$) and duty cycle (DC), as compared to quasars,
specially their more abundant subset, the
RQQs~\citep[e.g.,][]{GopalKrishna2003ApJ...586L..25G,Carini2003AJ....125.1811C,
  2004MNRAS.350..175S, Gupta2005A&A...440..855G,
  Carini2007AJ....133..303C, Goyal2012A&A...544A..37G}. It is thus
evident that INOV behaviour can be a powerful discriminator between
blazars and other powerful AGN, both radio-loud and
radio-quiet~\citep[e.g,][]{Carini2003AJ....125.1811C,2004MNRAS.350..175S,Goyal2012A&A...544A..37G,Goyal2013MNRAS.435.1300G}.
This point is discussed in Paper I and also in Sect. $4$ below.

To pursue the above clue, we extracted from the literature a
well-defined sample of 18 RQWLQs suited for our intranight optical
monitoring (Paper I). The sample was derived from the list of 86
radio-quiet WLQs published in Table 6 of~\citet[]{Plotkin2010AJ....139..390P}, based
on the SDSS Data Release 7~\citep[DR-7,][]{Abazajian2009ApJS..182..543A}. Out of that
list, we included in our sample all 18 objects brighter than R$\sim$18.5
which are classified as `high-confidence' BL Lac candidate based
on their optical spectra. INOV observations of 8 of the 18 RQWLQs were
reported in Paper I; these were carried out in $13$ sessions mainly with
the 130-cm Devasthal Fast Optical Telescope (DFOT) of the Aryabhatta
Research Institute of observational sciencES (ARIES). As part of the same continuing
program, we report here $16$ sessions of INOV observations of 10 RQWLQs with 
DFOT.

This paper is organized as follows. Section $2$ describes the
observations and data reduction, while Section $3$ gives details of
our statistical analysis. A brief discussion of our results is presented 
in Section $4$.


\section{Data Sample and observations} 

The present set of 10 RQWLQs listed in Table~\ref{tab:source_info},
was derived from the parent sample of $18$ RQWLQs mentioned above. The
selection threshold criteria of R $\sim$ 18.5 was adopted for the
sample so that 1-2m class telescopes would enable us to obtain a good
enough signal-to-noise ratio (SNR) for detecting fluctuations of $\sim
0.05$ mag with a reasonably good time resolution of $\sim 10$ minutes,
or better. In Paper I, we reported 13 sessions of intranight
monitoring of $8$ of these 18 RQWLQs. Here we have enlarged this study
by monitoring 10 RQWLQs (Table~\ref{tab:source_info}) in 16 sessions;
among them 7 RQWLQs being newly observed and three are repeated from
our sample in Paper I (namely, J081250.79$+$522531.05,
J084424.20$+$124546.00, J121929.50$+$471522.00). This has led to an
enlarged dataset of $29$ sessions covering $15$ RQWLQs, as discussed
in Sect. 4.

\begin{table}
\centering
\begin{minipage}{500mm}
{\scriptsize
\caption{The 10 RQWLQs studied in the present work. \label{tab:source_info}}

\begin{tabular}{lcc cc}
\hline
\multicolumn{1}{l}{IAU Name{\footnote{Sources marked by $^{*}$ were also reported in Paper I.}}} &  R.A.(J2000) & Dec(J2000)                       &{\it R} &   $z$  \\
         & (h m s)      &($ ^{\circ}$ $ ^{\prime}$ $ ^{\prime\prime }$) & (mag) &     \\
 (1)     &(2)             &(3)                             &(4)     &(5)  \\
\hline
\multicolumn{5}{l}{}\\

J081250.79$+$522531.05$^{*}$ &  08 12 50.80& $+$52 25 31 & 18.30 &1.152 \\
J084424.20$+$124546.00$^{*}$ & 08 44 24.20& $+$12 45 46 & 18.28 &2.466 \\
J090843.25$+$285229.80 &   09 08 43.25& $+$28 52 29 & 18.55 &0.930 \\
J101353.45$+$492757.99 &   10 13 53.45& $+$49 27 57 & 18.40 &1.635 \\ 
J110938.50$+$373611.60 &   11 09 38.50& $+$37 36 11 & 18.72 &0.397 \\
J111401.31$+$222211.50 &   11 14 01.31& $+$22 22 11 & 18.77 &2.121 \\
J115637.02$+$184856.50 &   11 56 37.02& $+$18 48 56 & 18.42 &1.956 \\
J121929.50$+$471522.00$^{*}$ & 12 19 29.50& $+$47 15 22 & 17.66 &1.336 \\
J212416.05$-$074129.90 &   21 24 16.05& $+$07 41 29 & 18.29 &1.402 \\
J224749.56$+$134250.00 &   22 47 49.56& $+$13 42 50 & 18.53 &1.179 \\
\hline
\end{tabular}

}
 \end{minipage}
\end{table}

\subsection{Photometric Observations}
Intranight monitoring for $\ga$ 3.5 hours of each RQWLQs
was carried out using the 130-cm DFOT of ARIES, located at Devasthal, 
India~\citep{Sagar2011Csi...101...8.25}.
It is a fast beam (f/4) optical telescope with a pointing accuracy 
better than $10$ arcsec RMS. DFOT is equipped with a 2K $\times$ 2K 
Peltier-cooled Andor CCD camera having a pixel size of
13.5 micron and with a plate scale of 0.54 arcsec per pixel. The CCD
covers a field of view of 18 arcmin on the sky. The CCD is
read out with $31$ and $1000$ kHz speeds, with  the corresponding 
system RMS noise of $2.5$, $7$ e- and gain of $0.7$, $2$ e-/Analog 
to Digital Unit (ADU). The CCD used in our observation
was cooled thermo-electrically to $-$85 degC. We observed 
each science frame for about $5-7$ minute, to achieve typical 
SNR better than $25-30$. The typical seeing FWHM during our monitoring sessions
was 2-2.5 arcsec.

In our sample selection, care was taken to ensure the availability of
at least two, but usually more, comparison stars covered within the CCD frame 
and also within $\sim$1 mag of the target RQWLQ. This allowed a reliable 
differential photometry by identifying and discounting any comparison 
star(s) found to vary during the monitoring session.

\subsection{Data Reduction}
\label{wl:sec_data}

The raw photometric data were processed using the standard tasks in the
Image Reduction and Analysis Facility {\textsc IRAF} \footnote{\textsc
  {Image Reduction and Analysis Facility (http://iraf.noao.edu/) }}.
For pre-processing of an image, we generated a master bias frame by
taking the median of all the bias frames and then subtracted it from
all the flat and source images. Master flat was generated by taking
the median of all the flat frames and then normalising the master flat.
Each source image was then flat-fielded by dividing by the normalised
master flat, in order to remove pixel-to-pixel inhomogeneities.
Finally, cosmic-ray removal was carried out from all source frames
using the IRAF task {\it cosmicrays}. The instrumental magnitudes of
the target and the comparison stars in the image frames were determined by
aperture photometry ~\citep{1992ASPC...25..297S, 1987PASP...99..191S}
and using the Dominion Astronomical Observatory Photometry software
\textrm{II} (DAOPHOT II). For the aperture photometry, we used four
aperture radii, $1\times$FWHM, $2 \times$ FWHM, $3 \times$ FWHM and $4
\times$ FWHM. Seeing disk radius (=FWHM/2) for each CCD frame was
determined by taking the mean value found for $5$ fairly bright unsaturated 
stars in each frame. The data reduced with different aperture radii were
found to be in good agreement. But the best SNR was almost always
found with aperture radius of $2 \times$ FWHM, so we adopted it for our
final analysis.

To derive the Differential Light Curves (DLCs) of a given RQWLQ monitored in a session, we 
selected two steady comparison stars present within the CCD
frames, on the basis of their proximity to the target RQWLQ,
both in location and magnitude. Coordinates of the comparison star pair
selected for each RQWLQ are given in Table ~\ref{tab_cdq_comp}. The
$g-r$ color difference for our `quasar-star' and `star-star' pairs
is always $< 1.6$ mag, with a median value of 0.7 (column 7,
Table ~\ref{tab_cdq_comp}). Detailed analyses by
~\citet{Carini1992AJ....104...15C} and
~\citet{2004MNRAS.350..175S} show that a color difference of this
magnitude should produce negligible effect on the DLCs as the
atmospheric attenuation changes during a monitoring session. \par

\begin{figure*}
\centering
\epsfig{figure=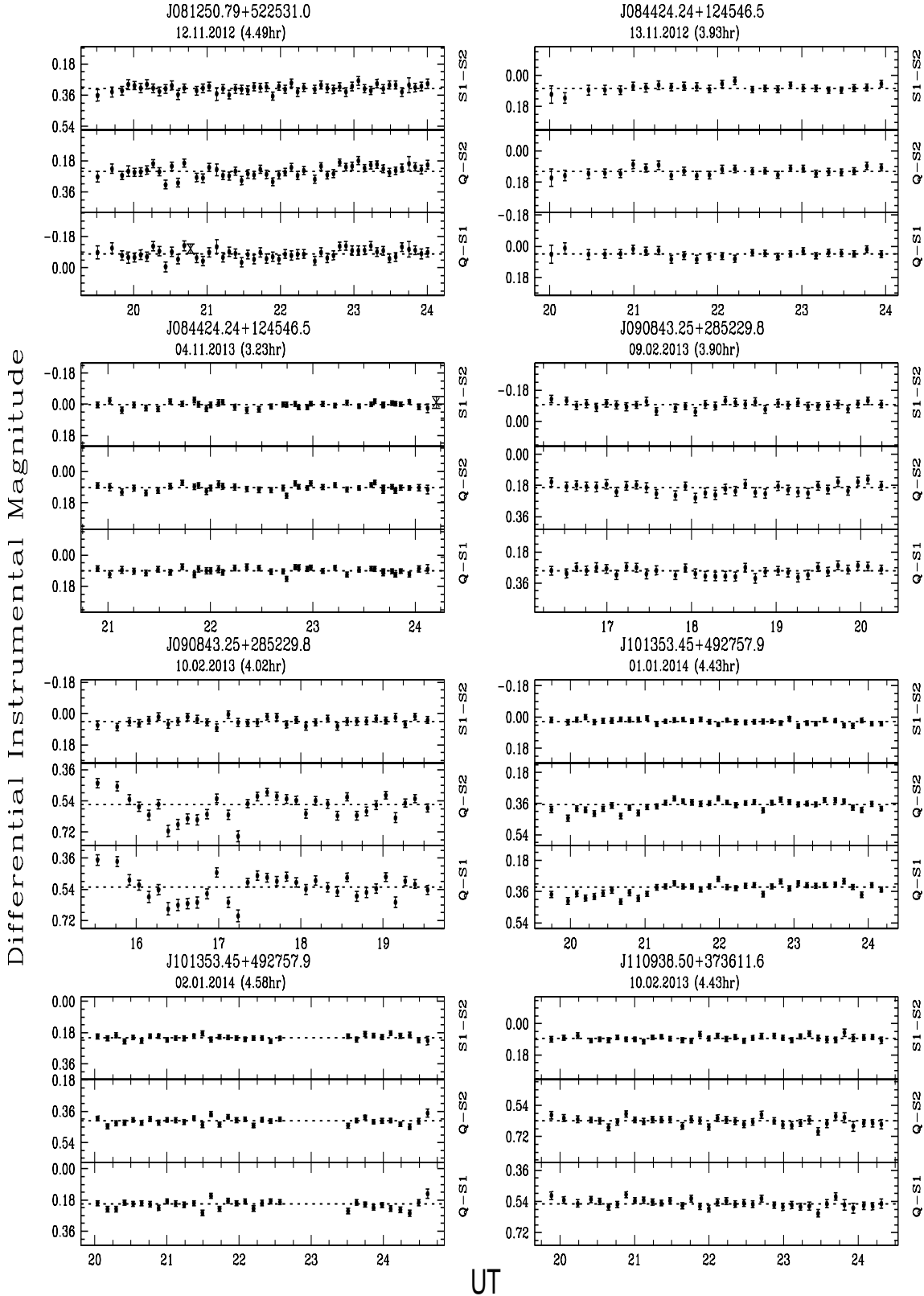,height=23.5cm,width=16.0cm,angle=00,bbllx=20bp,bblly=161bp,bburx=580bp,bbury=711bp,clip=true}
\vspace{-1.0cm}
\caption[]{Differential light curves (DLCs), for the $10$ RQWLQs in our sample. 
The name of the quasar along with the date and duration of the monitoring 
 session are given at the top of each panel. In each panel the upper DLC is 
 derived using the two non-varying comparison stars, while the lower two DLCs are the 
 `quasar-star' DLCs, as defined in the labels on the right side. Any likely 
 outlier point (at $> 3\sigma$) in the DLCs are marked with crosses 
 and those points are excluded from the statistical analysis.}
\label{fig:lurve}
 \end{figure*}

\begin{figure*} 
\centering
\epsfig{figure=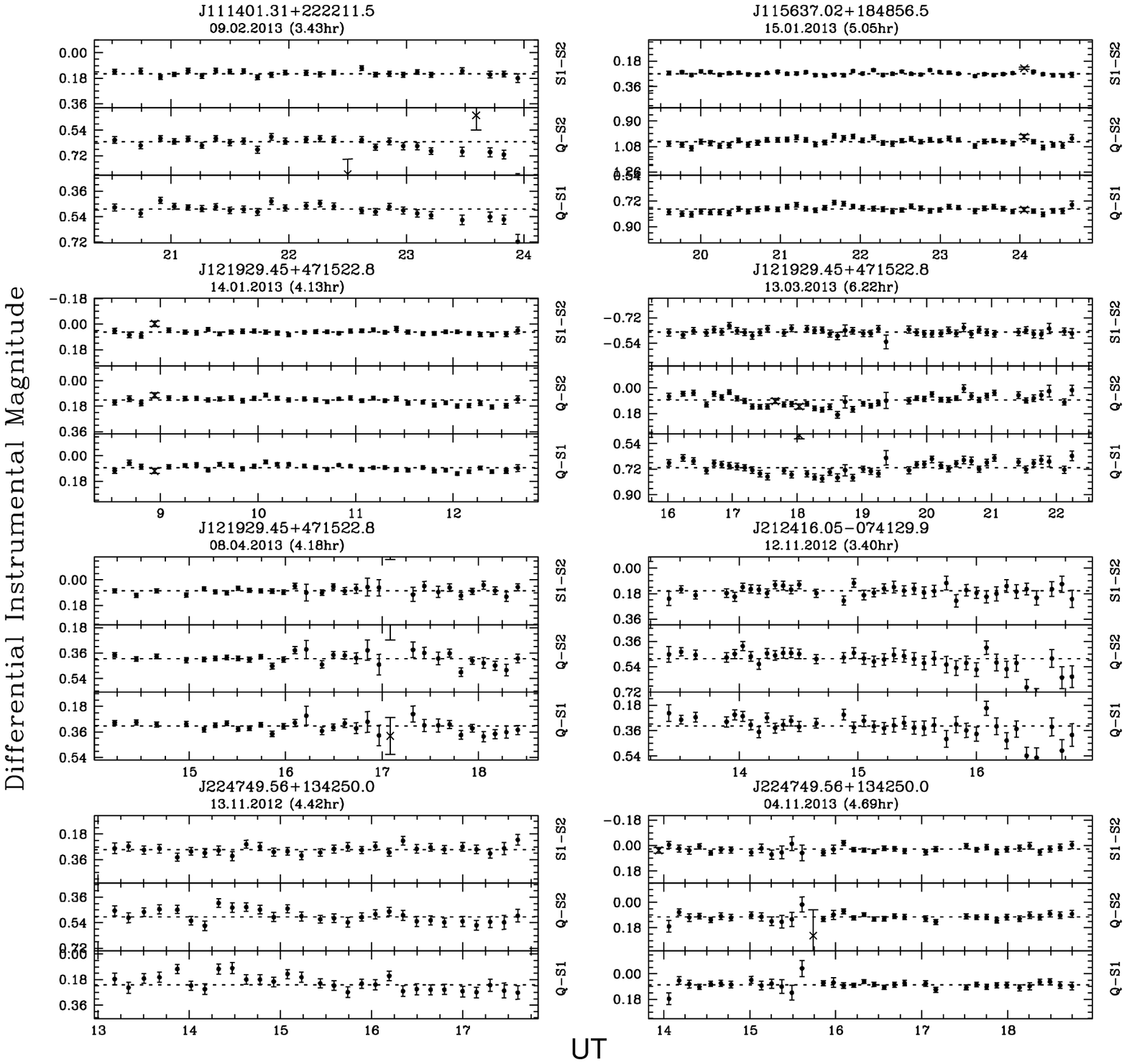,height=24.8cm,width=16.0cm,angle=00,bbllx=20bp,bblly=161bp,bburx=580bp,bbury=711bp,clip=true}
\vspace{-1.0cm}
\caption[]{Same as Figure~\ref{fig:lurve}, for remaining $8$ DLCs.}
\label{fig:lurve2}
 \end{figure*}

Since the selected comparison stars are non-varying, as judged from
the steadiness of their DLCs, any sharp fluctuation over a single temporal
bin was taken to arise due to improper removal of cosmic rays, or some
unknown instrumental effect,  and such outlier data points (deviating by
more than 3$\sigma$ from the mean) were removed from the affected DLCs,
by applying a mean clip algorithm.
In practice, such outliers were quite rare and never exceeded two data
points for any DLCs, as displayed in Figures \ref{fig:lurve},\ref{fig:lurve2}.

\begin{table*}
\centering
\caption{Basic parameters and observing dates of the 10 RQWLQs (and their
comparison stars).
\label{tab_cdq_comp}}
\scriptsize
\begin{tabular}{ccc ccc c}\\
\hline

{IAU Name} &   Date       &   {R.A.(J2000)} & {Dec.(J2000)}                      & {\it g} & {\it r} & {\it g-r} \\
           &  dd.mm.yy    &   (h m s)       &($^\circ$ $^\prime$ $^{\prime\prime}$)   & (mag)   & (mag)   & (mag)     \\
{(1)}      & {(2)}        & {(3)}           & {(4)}                              & {(5)}   & {(6)}   & {(7)}     \\
\hline
\multicolumn{7}{l}{}\\

J081250.79$+$522531.0 &  12.11.2012      &08 12 50.79 &$+$52 25 31.0  &   18.30 &       18.05 &        0.25\\   
S1                    &                  &08 13 28.04 &$+$52 19 33.2  &   19.57 &       18.19 &        1.38\\
S2                    &                  &08 13 52.52 &$+$52 27 01.0  &   18.99 &       17.79 &        1.20\\
J084424.24$+$124546.5 &  13.11.2012      &08 44 24.24 &$+$12 45 46.5  &   18.29 &       17.91 &        0.37\\
S1                    &                  &08 43 58.88 &$+$12 45 21.3  &   18.72 &       17.94 &        0.78\\
S2                    &                  &08 44 49.65 &$+$12 52 13.3  &   18.54 &       17.88 &        0.66\\
J084424.24$+$124546.5 &  04.11.2013      &08 44 24.24 &$+$12 45 46.5  &   18.29 &       17.91 &        0.37\\
S1                    &                  &08 44 17.37 &$+$12 50 18.3  &   19.30 &       17.89 &        1.41\\
S2                    &                  &08 44 39.26 &$+$12 44 54.7  &   18.25 &       17.88 &        0.37\\
J090843.25$+$285229.8 &  09.02.2013      &09 08 43.25 &$+$28 52 29.8  &   18.55 &       18.50 &        0.05\\
S1                    &                  &09 09 00.06 &$+$28 56 48.4  &   19.23 &       18.24 &        0.99\\
S2                    &                  &09 08 23.97 &$+$28 59 27.0  &   19.82 &       18.35 &        1.47\\
J090843.25$+$285229.8 &  10.02.2013      &09 08 43.25 &$+$28 52 29.8  &   18.55 &       18.50 &        0.05\\
S1                    &                  &09 08 58.83 &$+$28 55 38.9  &   18.92 &       17.93 &        0.99\\
S2                    &                  &09 08 27.19 &$+$28 52 20.3  &   18.16 &       17.89 &        0.27\\
J101353.45$+$492757.9 &  01.01.2014      &10 13 53.45 &$+$49 27 57.9  &   18.60 &       18.40 &        0.20\\
S1                    &                  &10 14 52.86 &$+$49 26 02.5  &   19.22 &       18.17 &        1.05\\
S2                    &                  &10 13 23.92 &$+$49 19 50.2  &   18.82 &       18.07 &        0.75\\
J101353.45$+$492757.9 &  02.01.2014      &10 13 53.45 &$+$49 27 57.9  &   18.60 &       18.40 &        0.20\\
S1                    &                  &10 13 58.83 &$+$49 32 13.2  &   19.36 &       18.24 &        1.12\\
S2                    &                  &10 14 22.84 &$+$49 29 08.9  &   18.69 &       17.98 &        0.71\\
J110938.50$+$373611.6 &  10.02.2013      &11 09 38.50 &$+$37 36 11.6  &   18.72 &       18.37 &        0.35\\
S1                    &                  &11 09 33.03 &$+$37 32 04.3  &   18.50 &       17.75 &        0.75\\
S2                    &                  &11 09 42.48 &$+$37 33 31.8  &   18.36 &       17.66 &        0.70\\
J111401.31$+$222211.5 &  09.02.2013      &11 14 01.31 &$+$22 22 11.5  &   18.77 &       18.38 &        0.39\\
S1                    &                  &11 14 20.88 &$+$22 29 41.2  &   19.30 &       18.01 &        1.29\\
S2                    &                  &11 14 32.46 &$+$22 17 36.1  &   19.28 &       17.87 &        1.41\\
J115637.02$+$184856.5 &  15.01.2013      &11 56 37.02 &$+$18 48 56.5  &   18.42 &       18.19 &        0.23\\
S1                    &                  &11 56 17.79 &$+$18 56 44.5  &   17.82 &       17.31 &        0.51\\
S2                    &                  &11 56 02.38 &$+$18 52 47.0  &   17.67 &       16.93 &        0.74\\
J121929.45$+$471522.8 &  14.01.2013      &12 19 29.45 &$+$47 15 22.8  &   17.65 &       17.53 &        0.12\\
S1                    &                  &12 19 06.03 &$+$47 13 10.8  &   19.00 &       17.55 &        1.45\\
S2                    &                  &12 19 22.97 &$+$47 09 31.0  &   17.93 &       17.44 &        0.49\\
J121929.45$+$471522.8 &  13.03.2013      &12 19 29.45 &$+$47 15 22.8  &   17.65 &       17.53 &        0.12\\
S1                    &                  &12 19 04.06 &$+$47 15 03.1  &   16.62 &       15.37 &        1.25\\
S2                    &                  &12 19 22.97 &$+$47 09 11.0  &   17.93 &       17.44 &        0.49\\
J121929.45$+$471522.8 &  08.04.2013      &12 19 29.45 &$+$47 15 22.8  &   17.65 &       17.53 &        0.12\\
S1                    &                  &12 20 17.33 &$+$47 18 04.4  &   17.63 &       17.23 &        0.40\\
S2                    &                  &12 19 05.60 &$+$47 07 36.4  &   18.61 &       17.24 &        1.37\\
J212416.05$-$074129.9 &  12.11.2012      &21 24 16.05 &$-$07 41 29.9  &   18.29 &       18.02 &        0.27\\
S1                    &                  &21 24 06.74 &$-$07 45 25.2  &   19.40 &       17.83 &        1.57\\
S2                    &                  &21 24 36.85 &$-$07 33 09.4  &   19.29 &       17.78 &        1.51\\
J224749.56$+$134250   &  13.11.2012      &22 47 49.56 &$+$13 42 50.0  &   18.53 &       18.26 &        0.27\\
S1                    &                  &22 47 49.24 &$+$13 47 04.3  &   19.44 &       18.30 &        1.14\\
S2                    &                  &22 48 16.29 &$+$13 39 05.8  &   19.75 &       18.37 &        1.38\\
J224749.56$+$134250   &  04.11.2013      &22 47 49.56 &$+$13 42 50.0  &   18.53 &       18.26 &        0.27\\
S1                    &                  &22 47 27.73 &$+$13 48 12.0  &   19.71 &       18.36 &        1.35\\
S2                    &                  &22 48 17.81 &$+$13 36 14.3  &   19.82 &       18.37 &        1.45\\

\hline
\end{tabular}
\end{table*}


\section{STATISTICAL ANALYSIS} 

Until a few years ago the most commonly used criterion  for testing the
presence of INOV is based on the so-called C-statistics, which is
defined as the ratio of standard deviations of the QSO-star DLC and
the corresponding star-star DLC~\citep[e.g.,][]{1997AJ....114..565J}.
Recently,~\citet{Diego2010AJ....139.1269D} has emphasised that the
usual definition of C-test is not a proper statistic, being based
on the ratio of standard deviations which (unlike the ratio of
variances) is not a linear statistical operator. They advocated more
powerful statistical tests, namely, the one-way analysis of variance
(ANOVA) and the F-test. However, a proper use of the ANOVA test
requires a rather large number of data points in the DLC, so as to
have several points within each subgroup used for the analysis. This
is not feasible for our DLCs since they typically have only around 25
- 50 data points each. Therefore, in this study we shall rely on the
F-test which is based on the ratio of variances, F $=
variance(observed)/variance(expected)$~\citep{Diego2010AJ....139.1269D}.
There are two versions of this test in the literature : (i) the standard
\emph{F$-$test}
~\citep[hereafter, $F^{\eta}-$test,][]{Goyal2012A&A...544A..37G} and
(ii) scaled \emph{F$-$test}
~\citep[hereafter, $F^{\kappa}-$test,][]{Joshi2011MNRAS.412.2717J}. The
latter test is more relevant in the cases where magnitudes of the
comparison stars are quite different from that of the target AGN
~\citep{Joshi2011MNRAS.412.2717J}. However for all our RQWLQs, we
could find comparison star within one magnitude of the RQWLQ
monitored. Therefore, we have applied the $F^{\eta}-$test to our DLCs.
Additionally, in view of recent detailed work by~\citet[][hereafter
  GGWSS13]{Goyal2013MNRAS.435.1300G} the $F^{\eta}-$test offers an
additional advantage, since our INOV results for RQWLQs can be
directly compared with the results published recently by GGWSS13 based
on the application of the $F^{\eta}-$test to the INOV observations of
several prominent classes of powerful AGN, taken in 262 monitoring
sessions.

Before applying the $F^{\eta}-$test, we recall here that the photometric 
errors, as returned by the routines in the data reduction softwares
(IRAF and DAOPHOT), are normally underestimated by a factor $\eta$ ranging 
between $1.3$ and $1.75$, as estimated in different studies~\citep[e.g.,][]
{1995MNRAS.274..701G, 1999MNRAS.309..803G, Sagar2004MNRAS.348..176S, 
Stalin2004JApA...25....1S, Bachev2005MNRAS.358..774B}.
In a recent analysis by~~\citet{Goyal2012A&A...544A..37G}, the best-fit value
of $\eta$ was estimated to be $1.5$. Following them, $F^{\eta}-$test can be expressed as :
\begin{equation} 
 \label{eq.fetest}
F_{1}^{\eta} = \frac{var(q-s1)}
{ \eta^2 \langle \sigma_{q-s1}^2 \rangle}, \nonumber  \\
\hspace{0.00cm} F_{2}^{\eta} = \frac{var(q-s2)}
{ \eta^2 \langle \sigma_{q-s2}^2 \rangle},\nonumber  \\
\hspace{0.00cm} F_{s1-s2}^{\eta} = \frac{var(s1-s2)}
{ \eta^2 \langle \sigma_{s1-s2}^2 \rangle}
\end{equation}
where $var(q-s1)$, $var(q-s2)$ and
$var(s1-s2)$ are the variances of the `quasar-star1',
`quasar-star2' and `star1-star2' DLCs and $\langle \sigma_{q-s1}^2
\rangle=\sum_\mathbf{i=1}^{N}\sigma^2_{i,err}(q-s1)/N$, $\langle \sigma_{q-s2}^2 \rangle$ and
$\langle \sigma_{s1-s2}^2 \rangle$ are the mean square (formal) rms
errors of the individual data points in the `quasar-star1', `quasar-star2'
and `star1-star2' DLCs, respectively. The scaling factor $\eta=1.5$, as mentioned above.

The $F^{\eta}$-test is applied by computing the $F$ values for individual DLCs, using
Eq.~\ref{eq.fetest}, and then comparing each computed value with the
critical $F$ value, $F^{(\alpha)}_{\nu_{qs},\nu_{qs}}$, where $\alpha$
is the significance level set for the test, and $\nu_{qs}$ is the
degrees of freedom of the given `quasar-star' DLC. Here, we used two
significance levels, $\alpha=$ 0.01 and 0.05, which correspond to
confidence levels of greater than 99 and 95 per cent, respectively. If $F$ is found
to exceed the critical value, the null hypothesis (i.e., no
variability) is discarded at the corresponding level of confidence.
Thus, a RQWLQ is marked as \emph{variable} (`V'), if both its DLCs (relative to the two 
comparison stars) have \emph{F-value} $\ge F_{c}(0.99)$, which corresponds to a confidence level
$\ge 99$ per cent, and \emph{non- variable} (`NV') if any one of its
two DLCs is found to have \emph{F-value} $\le F_{c}(0.95)$. Any remaining
cases are classified as \emph{probably variable} (`PV').

The inferred INOV classification for the DLCs of each RQWLQ, drawn
relative to two comparison stars, is presented in
Table~\ref{wl:tab_res}. In the first 4 columns, we list the name of
the RQWLQ, date of its monitoring, duration of monitoring and the
number of data points (N) in the DLCs. The next two column give the
computed F-values and the INOV status of the two DLCs of the RQWLQ,
as inferred from the $F^{\eta}-$test. Column $7$ gives the
session-averaged photometric error $\sigma_{i,err}(q-s)$ for the
`quasar$-$star' DLCs (i.e., mean value for q-s1 and q-s2 DLCs).
Typically, it lies between 0.02 and 0.07 mag (without the $\eta$ scaling
mentioned above). The last column gives the peak-to-peak INOV
amplitude $\psi$ based on the definition given
by~\citet*{Romero1999A&AS..135..477R}, as

\begin{equation} 
\psi= \sqrt{({D_{max}}-{D_{min}})^2-2\sigma^2} 
\end{equation} 

with  $D_{min,max}$ = minimum (maximum) of values observed in the RQWLQ DLC and $\sigma^2$=
$\eta^2$$\langle\sigma^2_{q-s}\rangle$, where,
$\eta$ is taken to be 1.5~\citep{Goyal2012A&A...544A..37G}. \par

The INOV duty cycle (DC) for our RQWLQ sample was computed
using the definition of~\citet*{Romero1999A&AS..135..477R}, as
\begin{equation} 
DC = 100\frac{\sum_{i=1}^n N_i(1/\Delta t_i)}{\sum_{i=1}^n (1/\Delta t_i)} 
{\rm per cent} 
\label{eqno1} 
\end{equation} 
where $\Delta t_i = \Delta t_{i,obs}(1+z)^{-1}$ is duration of the
monitoring session of a RQWLQ on the $i^{th}$ night, corrected for
its cosmological redshift, $z$. Since the duration of the observing
session for a given RQWLQ differs from night to night,
the computation of DC has been weighted by the actual monitoring
duration $\Delta t_i$ on the $i^{th}$ night. $N_i$ was set equal to 1,
if INOV was detected (i.e., `V'), otherwise $N_i$ was taken as $0$.


\section{RESULTS AND DISCUSSION}
\label{wl:sec_dis}

This paper extends our work of Paper I which reported the first
systematic investigation of the INOV properties of radio-quiet
weak-line quasars (RQWLQs). To the 13 DLCs reported in Paper I, we
have added here 16 DLCs of durations $\ga 3.5$ hours
(Table~\ref{wl:tab_res}), derived for 10 RQWLQs of which three were
also included in Paper I. Table 3 presents our INOV results for the 10
RQWLQs. These are based on the $F^{\eta}$-test (Eq. 1) which is a more
reliable version of the $F$-test
~\citep{Howell1988AJ.....95..247H,Diego2010AJ....139.1269D}, as shown
by GGWSS13 who applied it to determine the INOV status of 262 DLCs of
77 AGN representing 6 prominent classes of AGN (see below). These
authors adopted $\eta = 1.5$, as determined
by~\citet{Goyal2012A&A...544A..37G} from their analysis of a large set
of 262 DLCs of comparison stars. It was also shown by GGWSS13 that the
INOV duty cycles determined using the $F^{\eta}$-test are
indistinguishable from those found using the `modified C-test', again taking
$\eta = 1.5$. The $F^{\eta}$-test applied here to the 16 DLCs of 10
RQWLQs, taking $\eta = 1.5$, yielded an INOV duty cycle of 5\% which
rises to 15\% if the single case of `probable' INOV (PV) is included
(Table~\ref{wl:tab_res}). The same result is obtained using the
`modified C-test', consistent with the finding by GGWSS13, as
mentioned above. In order to ascertain the effect of likely
uncertainty in $\eta$ value, we have repeated the computation of INOV
duty cycle for the 10 RQWLQs, taking two extreme values for $\eta$ (=1.3 and 1.75)
reported in the literature~\citep[][and references
  therein]{Goyal2012A&A...544A..37G}. The INOV duty cycle computed
using these extreme values of $\eta$ range up to 15\% which can be
treated as an upper limit.\par

Next, we have computed the INOV duty cycle for the enlarged sample of
29 DLCs obtained by combining our present INOV observations of $10$
RQWLQs with those reported in Paper I. We thus find the INOV duty
cycle for the combined set of $15$ RQWLQs to be $\sim$5\%, rising to
$\sim$11\% if the DLCs classified as `probable' INOV (PV) are
included. It is interesting to compare these estimates found here for
RQWLQs with those reported by GGWSS13 for several other AGN classes,
following an essentially identical observing and analysis procedure.
The INOV duty cycle inferred by them (using the $F^{\eta}$-test taking
$\eta$ = 1.5) is ~ 10\%(6\%) for radio-quiet quasars (RQQs), ~
18\%(11\%) for radio-intermediate quasars (RIQs), ~ 5\%(3\%) for radio
lobe-dominated quasars (LDQs), ~17\%(10\%) for radio core-dominated
quasars with low optical polarization (LPCDQs) , ~43\%(38\%) for radio
core-dominated quasars with high optical polarization (HPCDQs) and
~45\%(32\%) for BL Lac objects (BLOs) (The values inside parentheses
refer to the DLCs showing INOV amplitude $\psi > 3$\%). The INOV duty cycle for Seyfert galaxies is reported to
lie between 10\% and 20\%, the higher values being associated with the
radio-loud subset~~\citep[e.g., see][]{Carini2003AJ....125.1811C}. Finally, we note that the
apparent similarity of the DC estimates found here for
the RQWLQs with the afore-mentioned estimates given in GGWSS13 for RQQs,
RIQs, LDQs and LPCDQs is likely to be superficial. This is because, in contrast
to the INOV detection threshold, $\psi_{lim}$, of 1-2\% characteristic
of the observations used in GGWSS13, $\psi_{lim}$ reached in our INOV
programme for the RQWLQs is a factor 2-3 higher, essentially because
the RQWLQs are typically 1-2 mag fainter compared to the AGN samples
covered in GGWSS13. Therefore, the present estimates of INOV duty cycle
for the RQWLQs may well have to be revised upwards. A proper comparison has to wait for the
availability of about a magnitude more sensitive INOV observations for
RQWLQs, compared to those reported here and in Paper I. Such
sensitivity matched INOV observations of RQWLQs may well yield
substantially higher INOV duty cycles than those estimated here,
perhaps approaching the values obtained for HPCDQs or BLOs. Efforts
are underway to use larger telescopes for intranight monitoring of
RQWLQs. \par

As of now, our programme has revealed two instances of RQWLQ
exhibiting an INOV amplitude $\psi >$ 3\% in a monitoring session, a
level rarely observed in our 2-decade long INOV
programme (summarised in GGWSS13), except for
BLOs and HPCDQs. The two RQWLQs,
J090843.25$+$285229.8 ($\psi \sim$ 31\% on 10-02-2013, Table 3) and
J121929.45$+$471522.8 ($\psi \sim$ 7\% on 26-02-2012, Paper I), are
thus the best available candidates for the elusive radio-quiet BLOs
and both need to be followed up. Further INOV observations of these
and several other members of our sample of 18 relatively bright RQWLQs
are planned for the next winter months.

\begin{table*}
 \centering
 \begin{minipage}{500mm}
 {\small
 \caption{Observational details and INOV results for the sample of 10 RQWLQs over 16 monitoring sessions.}
 \label{wl:tab_res}
 \begin{tabular}{@{}ccc cc rrr rrr ccc@{}}
 \hline  \multicolumn{1}{c}{RQWLQ} &{Date} &{T} &{N} 
 &\multicolumn{1}{c}{F-test values} 
 &\multicolumn{1}{c}{INOV status{\footnote{V=variable, i.e., confidence level
       $\ge 0.99$; PV=probable variable, i.e., $0.95-0.99$ confidence level;
       NV=non-variable, i.e., confidence level $< 0.95$.\\
 Variability status values based on quasar-star1 and quasar-star2
 pairs are separated by a comma.}}}
 &{$\sqrt { \langle \sigma^2_{i,err} \rangle}$}&{INOV amplitude}\\
 & dd.mm.yyyy& 
hr & &{$F_1^{\eta}$},{$F_2^{\eta}$}
 &F$_{\eta}$-test &(q-s) &$\psi_1(\%),\psi_2$(\%)&$\frac{}{}$\\
 (1)&(2) &(3) &(4) &(5)&(6)
 &(7) &(8)\\ \hline

J081250.79$+$522531.0  &12.11.2012 & 4.49&    50&    0.44,    0.78&       NV, NV&   0.04&   10.70,      12.91\\   
J084424.24$+$124546.5  &13.11.2012 & 3.93&    25&    0.23,    0.33&       NV, NV&   0.04&    3.91,       5.66\\
J084424.24$+$124546.5  &04.11.2013 & 3.23&    38&    0.45,    0.50&       NV, NV&   0.02&    6.15,       6.95\\
J090843.25$+$285229.8  &09.02.2013 & 3.90&    32&    0.33,    0.44&       NV, NV&   0.04&    5.06,       8.90\\
J090843.25$+$285229.8  &10.02.2013 & 4.02&    33&    3.01,    3.14&       V,  V&    0.04&   31.73,      30.20\\
J101353.45$+$492757.9  &01.01.2014 & 4.43&    37&    1.96,    1.58&       PV, NV&   0.02&   12.79,      11.07\\  
J101353.45$+$492757.9  &02.01.2014 & 4.58&    32&    1.10,    0.78&       NV, NV&   0.02&   10.68,       7.31\\  
J110938.50$+$373611.6  &10.02.2013 & 4.43&    36&    0.54,    0.52&       NV, NV&   0.03&    9.89,       9.14\\ 
J111401.31$+$222211.5  &09.02.2013 & 3.43&    25&    2.15,    2.80&       PV,  V&   0.04&   28.47,      30.43\\
J115637.02$+$184856.5  &15.01.2013 & 5.05&    41&    0.59,    0.74&       NV, NV&   0.03&    7.60,       7.97\\  
J121929.45$+$471522.8  &14.01.2013 & 4.13&    33&    0.87,    0.84&       NV, NV&   0.02&    7.54,       7.89\\ 
J121929.45$+$471522.8  &13.03.2013 & 6.22&    44&    1.30,    1.48&       NV, NV&   0.04&   15.16,      17.86\\
J121929.45$+$471522.8  &08.04.2013 & 4.18&    30&    0.50,    0.59&       NV, NV&   0.05&   14.01,      14.09\\
J212416.05$-$074129.9  &12.11.2012 & 3.40&    37&    1.08,    1.07&       NV, NV&   0.07&   33.33,      35.20\\
J224749.56$+$134250.0  &13.11.2012 & 4.42&    29&    0.89,    0.71&       NV, NV&   0.05&   15.30,      14.32\\ 
J224749.56$+$134250.0  &04.11.2013 & 4.69&    35&    0.63,    0.45&       NV, NV&   0.04&   20.45,      14.51\\ 
 \hline
 \end{tabular}
 }
 \end{minipage}
 \end{table*} 

\section*{Acknowledgments}
 We gratefully acknowledge assistance from the staff of the 130-cm
 DFOT telescope of ARIES Nainital and Ravi Joshi in resolving some
 issues related to data reduction.

\bibliography{references}
\label{lastpage}
\end{document}